\documentclass[pra,preprint,showpacs]{revtex4}
\usepackage[centertags]{amsmath}
\usepackage{amsfonts}
\usepackage{amssymb}
\usepackage{amsthm} 
\usepackage{newlfont}
\usepackage{epsfig}
\usepackage{amscd}
\usepackage{graphicx}
\usepackage{epsfig}
\usepackage{footnote}
\usepackage{lipsum}
\usepackage{color}
\usepackage{xcolor}
\usepackage{graphicx}
\usepackage{multirow}
\usepackage{subfig}
\usepackage{amscd}

\newcommand{\beq}{\begin{equation}}
\newcommand{\eeq}{\end{equation}}
\newcommand{\ba}{\begin{array}}
\newcommand{\ea}{\end{array}}
\newcommand{\bea}{\begin{eqnarray}}
\newcommand{\eea}{\end{eqnarray}}
\begin{document}

\begin{center}
{\large \sc \bf {Computing scalar products via a two-terminal quantum transmission line
}
}

\vskip 15pt

{\large 
J.Stolze$^1$ and A.I.~Zenchuk$^2$ 
}

\vskip 8pt

{\it $^1$Technische Universit\"at Dortmund, Fakult\"at Physik, D-44221 Dortmund, Germany},\\
{\it $^2$Institute of Problems of Chemical Physics, RAS,
Chernogolovka, Moscow reg., 142432, Russia}.

\vskip 8pt

\end{center}

\begin{abstract}
The scalar product of two vectors with $K$ real components
can be computed using two quantum channels, that is, information
transmission lines in the form of spin-1/2 XX chains.  Each channel
has its own $K$-qubit sender and both channels share a single
two-qubit receiver. The $K$ elements of each vector are encoded in the
pure single-excitation initial states of the senders. After time evolution,
a bi-linear combination of these elements appears in the only matrix
element of the second-order coherence matrix of the receiver state. An
appropriate local unitary transformation of the extended receiver
turns this combination into a renormalized version of the scalar
product of the original vectors.  The squared absolute value of this
scaled scalar product is the intensity of the second-order coherence
which consequently can be measured, for instance, employing
multiple-quantum NMR. The unitary transformation generating the scalar
product of two-element vectors is presented as an example.
\end{abstract}


\maketitle

\section{Introduction}
\label{Section:introduction}
The realization of  classical computational algorithms on the basis of quantum operations has become a promising direction in the development of quantum information processing. A  well known algorithm of this type is  the   HHL  (Harrow-Hassidim-Lloyd) algorithm for solving  systems of linear algebraic equations \cite{HHL}, which was realized for solving the simplest systems of two equations on the basis of photonic systems   \cite{CWSCGZLLP} and superconducting quantum gates \cite{ZSCetal}. Some other applications of this algorithm can be found in \cite{CJS,BWPRWL}. 

Among the  powerful and effective quantum algorithms included 
in most of 
the contemporary quantum counterparts of the classical algorithms of computational algebra (both linear and nonlinear) are the quantum Fourier transform (QFT) \cite{NCh}, Hamiltonian simulation \cite{BACS,Ch} and phase estimation \cite{LP,CEMM,NCh} (based on  QFT).
These subroutines are used in  the above mentioned HHL-algorithm  and in the   algorithms 
 for  solving  systems of nonlinear equations \cite{QHL},  {executing} various  elementary matrix operations (including addition, multiplication and {tensor} product) {with a method based on the Trotter product
formula} \cite{ZZRF},  performing the effective measurement of the desired observable \cite{PP} (which is  applicable to  various problems of linear algebra),
 and
  solving linear differential equations \cite{BCOW}. 
In all these quantum algorithms, the  phase estimation as an essential subroutine allows to increase the accuracy of calculations  using  an additional quantum subsystem whose dimensionality increases with an increase in the desired accuracy. 

We consider a quantum counterpart of a particular algebraic problem widely applicable in different fields of physics and mathematics, namely, the scalar product of arbitrary vectors {with real entries (the way to generalize it for complex vectors is also discussed)}. 
{ A  probabilistic protocol of scalar product estimation was  proposed in \cite{ZFF}  and then resumed in the Appendix B of \cite{ZZRF}. In that protocol, {both} vectors to be multiplied are encoded into the same system, they are multiplied by the states, respectively,  $|0\rangle$ and $|1\rangle$   of an additional qubit (ancilla).  Then, after applying the Hadamard operator (or phase rotation) to the ancilla,  one obtains the real (imaginary) part of the scalar product   as the probability of measuring $|0\rangle$ at the ancilla.  {The precision of the result depends on the number of repetitions of the protocol.}

 { In contrast, our protocol is a single-shot operation.}  The vectors to be multiplied are the initial  states of two different senders. Therefore, these states can be output states of other quantum algorithms. Senders can be remote   one from another, {so that time}  evolution is required to transfer sender's states to the receiver. Applying the proper unitary transformation to the extended receiver we obtain the   scalar product of  { the input} vectors as a  particular element (second-order coherence matrix) of the two-qubit receiver's density matrix. 
 {Being} localized in the particular matrix element, the scalar product can be used as  input data  for other quantum operations. Thus, the quantum schemes proposed in our paper can be blocks in a more general device, and  the protocol of scalar product can be used as a subroutine in other algorithms.
}


{The hardware used in} our algorithm is a two-channel communication line. Each channel includes the $K$-qubit sender, the transmission line and one qubit of the receiver. Thus, the receiver consists of two qubits {independently of} the dimensionality of the vectors to be multiplied. 
{However}, the number of qubits $K$ in each sender equals the
dimensionality of these vectors. {The  components of the vectors to be
  multiplied are the probability amplitudes of the senders' initial
  states, { therefore the norm of these vectors is bounded  by
    one}. These amplitudes are real in the case of real vector
  components.}  The transmission lines are necessary  to connect the
remote senders with  the receiver and can {consist of}  different
numbers of qubits.

{After evolution, the elements of the} vectors {${\mathbf v}^{(i)}$,
  $i=1,2$,} encoded  {into} the initial states of the
senders appear in a bi-linear combination in the ''corner'' element
(the element $\rho_{14}$ of the $4$-dimensional  receiver's density
matrix which is also the only element of the  
2nd-order MQ coherence matrix). {Then, using the unitary transformation of the extended receiver, we eliminate all those terms in that bilinear combination which do not contribute to the desired scalar product. In the end, the corner element of the density matrix is equal to the scalar product of the original vectors multiplied by a scalar factor.}
 We show that  this scalar factor  { equals $\displaystyle s \sqrt{(1-|{\mathbf v}^{(1)}|^2) (1-|{\mathbf v}^{(2)}|^2)}$} where  $s$ takes its maximal value 
 $\displaystyle s=\frac{1}{\sqrt{K}}$ in the shortest communication line of $2 K$ qubits (there are no transmission lines in this case)  and decreases with an increase in the length of the channels.   

\section{Scheme of communication line  and its evolution. }
\label{Section:SimpleSch}

\begin{figure*}
\epsfig{file=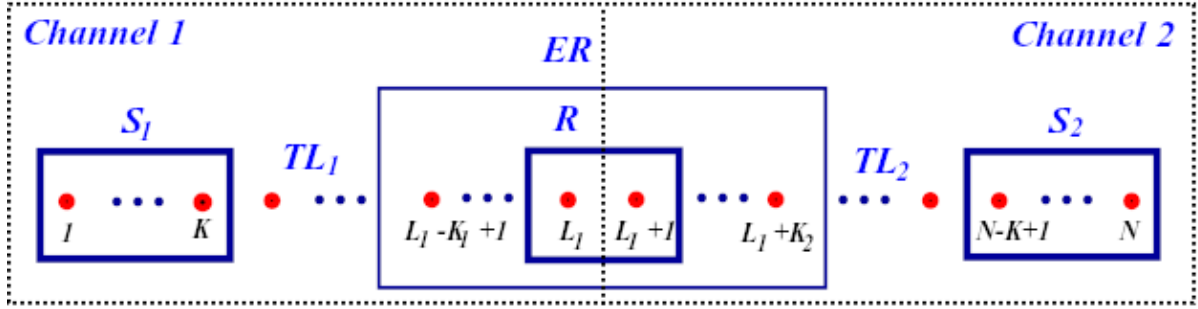,
  scale=3
   ,angle=0
}  
\caption{General scheme of the spin-1/2 two-channel communication line
  for performing the scalar product of remote $K$-element  vectors
  encoded  in  the pure states of the  $K$-qubit senders $S_1$ and
  $S_2$. The result of the  scalar product appears in the element
  $\rho^{(R)}_{14}$  of the 2-qubit
  receiver's density matrix. {The extended receiver $ER$ consists of
  $(K_1+K_2)$ qubits}}
  \label{Fig:CL1} 
\end{figure*}

The general 
 setup
 of  the two-channel  spin chain proposed for implementing the
 protocol of the scalar product of two $K$-element vectors initially
 remote from each other  is illustrated
  in Fig.\ref{Fig:CL1}. It consists of two $K$-qubit senders $S_1$ and $S_2$, which are used to encode the
elements of the vectors as is shown below, the two-qubit receiver $R$
for registration of the result, the $(K_1+K_2)$-qubit   extended
  receiver $ER$ (the values of $K_i$, $i=1,2$, will be 
 determined
 below) for applying the required unitary transformations, and two transmission lines  $TL_i$, $i=1,2$,  connecting the senders with the receiver. We emphasize that the lengths $L_1$ and $L_2$ of the first and second channels might be different (due to the different lengths of $TL_1$ and $TL_2$), and the  full  length of the communication line is  $N=L_1+L_2$. The receiver consists of two qubits, which are the end-nodes of the channels.
The extended receiver $ER$  encompasses  {$K_1$ and $K_2$ qubits from, respectively, the first and second channel.}

\begin{figure*}
\epsfig{file=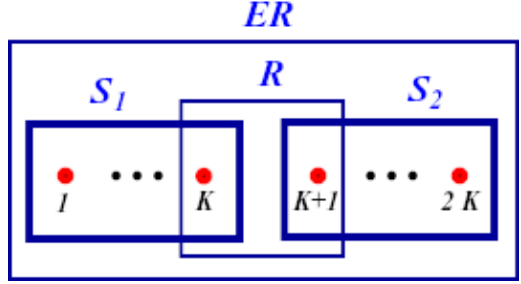,
  scale=3
   ,angle=0
}  
\caption{ The simplest $2K$-qubit spin-1/2 system
  allowing to perform the scalar product of $K$-element  vectors
  encoded in the pure states of the  $K$-qubit senders
  $S_1$ and $S_2$. The result of the  scalar product
  appears in the element $\rho^{(R)}_{14}$  of the 2-qubit receiver's density matrix. 
}
  \label{Fig:CL2} 
\end{figure*}

In the simplest case of the $2K$-spin chain ($N=2 K$) the scheme
reduces to the  one shown in Fig.\ref{Fig:CL2}. There is no $TL_i$,
$i=1,2$, in this scheme. In addition, $R$ overlaps with $S_1$ and
$S_2$, and the extended receiver $ER$ is
 identical to
 the  complete spin system. 

\subsection{Evolution of 
the
communication line and final state of
 the receiver.}
Let the evolution of  the spin chain be governed by the $XX$ Hamiltonian 
\begin{eqnarray}\label{XX}
&&H=\sum_{i=1}^{N-1} D_{i} (I_{i,x}I_{i+1,x} +I_{i,y}I_{i+1,y}),\\\label{comm0}
&&[H,I_z]=0.
\end{eqnarray}
The $I_{i,\alpha} (i=1,...,N ; \alpha=x,y,z)$ are spin-1/2 operators with eigenvalues $\pm 1/2$  and $I_z$ is the total $z$ component, $I_z=\sum_{i=1}^N I_{i,z}$.
We  also need a  unitary transformation $U^{(ER)}(\varphi)$ of the
extended receiver, where $\varphi$ denotes a list of free parameters
of this transformation. The conservation law
\begin{eqnarray}\label{constr}
[U^{(ER)},I^{(ER)}_z]=0
\end{eqnarray}
(where $I_z^{(ER)}$ is the total $z$ spin component of the extended receiver) {together with (\ref{comm0})}
prevents the mixing of matrix elements from the coherence matrices of different orders { \cite{FZ_2017}}.
After initialization, we first allow the spin chain to evolve till
some optimal time instant $t_0$ (which will be determined below for a
particular example, see Sec.\ref{Section:Example2}). Then, at time
$t_0$, we apply the unitary transformation to the  extended receiver. 
Thus, the complete unitary transformation of the communication line reads 
\begin{eqnarray}\label{W}
&&
W=\Big(E_{S_1,TL_1'}\otimes U^{(ER)}(\varphi)\otimes E_{TL_2',S_2} \Big)  V(t), \\\label{Vt}
&&
V(t)=e^{-i H t},
\end{eqnarray}
where $TL_i'$, $i=1,2$, are the transmission lines $TL_i$  without the nodes of the extended receiver and $E_{S_i,TL_i'}$ is the identity operator in the space of the system  $S_i \cup TL_i'$, $i=1,2$.

In the case shown in Fig.\ref{Fig:CL2}, the united senders, $S_1 \cup
S_2$, the extended receiver, and the complete system coincide. Hence,
it is not necessary to wait for the time evolution to transfer the
quantum information from the senders to the extended receiver. Therefore, we can disregard 
 time
 evolution and apply the 
unitary transformation $U^{(ER)}$ immediately (at $t=0$),
so that the operator $W$   reads  
\begin{eqnarray}
W=U^{(ER)}(\varphi) .
\end{eqnarray}
The receiver density matrix reads
\begin{eqnarray}\label{rhoR2}
\rho^{(R)}={\mbox{Tr}}_{\!/R} \;(\rho),
\end{eqnarray}
where the trace is calculated over all the nodes except
those of the receiver.  
{
In the following section, the ``senders only'' scheme from
Fig.\ref{Fig:CL2} will be discussed in more detail.

\section{Scalar product via  the  ``senders only'' scheme}
\label{Section:ScalarProdD2}

First, we consider the scalar product of $K$-element vectors using
 the scheme in Fig.\ref{Fig:CL2} 
where $N=2K$ so that there 
 are
 no transmission lines and even the
receiver nodes are contained in the senders.
We start with the 
  pure initial states of the senders $S_1$ and $S_2$ 
\begin{eqnarray}\label{pureK}
|\psi_i\rangle = a^{(i)}_0|0\rangle + \sum_{n=1}^K a^{(i)}_{{K-n+1}} |n\rangle,\;\;\sum_{n=0}^K|a^{(i)}_n|^2=1,\;\;i=1,2,\;\;{ a^{(i)}_0\neq 0}.
\end{eqnarray}
Here $|n\rangle$ means the one-excitation state with 
the $n$th spin of $S^{(1)}$ or $S^{(2)}$ excited.
The initial state of the 
 complete
system  reads
 \begin{eqnarray}\label{inst2}
\rho(0)=\rho^{(S_1)}(0)\otimes  \rho^{(S_2)}(0),\;\;
{\rho^{(S_i)} = |\psi_i\rangle \langle \psi_i|,\;\;i=1,2}.
\end{eqnarray}
This initial state has no more than two excitations, therefore the
system evolves in the  zero-, one- and two excitation state subspaces
due to the 
conservation laws
(\ref{comm0}) and (\ref{constr}).
 In this section, 
we will use capital Latin multi-indices, where 
the subscripts 1 and 2 are related, respectively, to
$S_1$ and $S_2$   (for instance, $I_1$, $I_2$), the multi-index with the subscript $R$ is related to the receiver ($I_R$), and primed multi-indices are related to the appropriate sender without the receiver's nodes (for instance, $I_1'$, $I_2'$). 
 
We write 
the receiver density matrix 
(\ref{rhoR2}) in components, explicitly performing the matrix multiplications and also the trace over all
degrees of freedom except those of the receiver:
\begin{eqnarray}
\label{eq:10}
\rho^{(R)}_{N_R;M_R}= \sum_{N_1',N_2',I_1,I_2,J_1,J_2}  W_{N_1'N_RN_2';I_1I_2} \rho^{(S_1)}_{I_1;J_1} \rho^{(S_2)}_{I_2;J_2} W^+_{J_1J_2;N_1' M_R N_2'} .
\end{eqnarray}
Since the receiver  is a 2-qubit system, its
coherence matrix of order +2 has only a single element, connecting the
two-qubit states with multi-indices
$N_R=\{0,0\}$ and $M_R=\{1,1\}$, which we now construct. Due to
(\ref{constr}) the matrix $W$ is block-diagonal with respect to the
number of excitations. Since $M_R=\{1,1\}$, the multi-indices $N_1'$
and $N_2'$ on $W^+$ can only contain zero entries, which we denote by
$0_1'$ and $0_2'$, respectively.  {Because of the one-excitation initial states (\ref{pureK}) of the senders, $J_1$ and $J_2$   must contain
exactly one entry equal to unity}, which we denote by
$|J_1|=|J_2|=1$. Similarly, since $N_R=\{0,0\}${,} and $N_1'=0_1'$ and
$N_2'=0_2'$, the matrix $W$ operates in the zero-excitation subspace
and hence has only one element  $W_{0_1'0_R0_2';0_10_2}=1$, where again
$0_1$ and $0_2$ denote the zero-excitation states of the senders $S_1$
and $S_2$, respectively. The desired element thus reads

\begin{eqnarray}\label{rhoR}
\rho^{(R)}_{00;11}= \sum_{{|J_1|=1}\atop{|J_2|=1}}\rho^{(S_1)}_{0_1;J_1} \rho^{(S_2)}_{0_2;J_2} W^+_{J_1J_2;0_1' 11 0_2'} .
\end{eqnarray}
  This equation involves only one column of $W^+$,  indexed by
  $(0'_1110_2')$.  
The other columns must fulfill the unitarity condition $W W^+= E_{2K}$
(the $2K \times 2K$ unit matrix) and they are arbitrary otherwise.

In order to connect to the pure initial states (\ref{pureK}) we rewrite
the element (\ref{rhoR}) as follows:
\begin{eqnarray}\label{rho0011}
&&\rho^{(R)}_{00;11} = \sum_{j=1}^K\sum_{i=1}^K s_{ij} \rho^{(S_1)}_{0_1;J_1^{(i)}} \rho^{(S_2)}_{0_1;J_2^{(j)}}, 
\\\label{s}
&&s_{ij}=W^+_{J^{(i)}_1J^{(j)}_2;0_1'110_2'} .
\end{eqnarray} 
The coefficients $s_{ij}$ are related to  the elements {(which are still free)} of the unitary transformation $W\equiv U^{(ER)}$ and $J^{(i)}_k$ 
is the multi-index $J_k$ with all zeros except the  $i$th entry which
equals 1. 
We set
 certain coefficients $s_{ij}$ equal to zero,  keeping in mind that
 the matrix $W$ has to be unitary:
\begin{eqnarray}\label{sconstr}
s_{ij}=\delta_{ij}s,\;\;i,j=1,\dots,K.
\end{eqnarray}
Then  (\ref{rho0011}) reduces to 
\begin{eqnarray}\label{sp}
&&\rho^{(R)}_{00;11}=s \sum_{i=1}^K \rho^{(S_1)}_{0_1;J_1^{(i)}} \rho^{(S_2)}_{0_1;J_2^{(i)}} =
{S \sum_{i=1}^K \Big(a^{(1)}_{i} a^{(2)}_{i}\Big)^*,}\\\label{S}
&&{S=a^{(1)}_{0} a^{(2)}_{0}  s},
\end{eqnarray}
since 
the elements of the  initial senders' density matrices are given by
\begin{eqnarray}\label{star}
\rho^{(S_k)}_{0_k;J_k^{(i)}} = a_0^{(k)} (a_i^{(k)})^*.
\end{eqnarray}
If the $a^{(j)}_{i}$ are all real, { we can introduce the vectors 
\begin{eqnarray}\label{vectors}
{\mathbf v}^{(j)}=\left(
\begin{array}{c}
a^{(j)}_{1}\cr
\cdots\cr
a^{(j)}_{K}
\end{array}
\right),\;\;j=1,2,
\end{eqnarray}
then 
the rhs of eq.(\ref{sp})
is proportional to the scalar product of these  vectors:
\begin{eqnarray}\label{vec}
\rho^{(R)}_{00;11}=S {\mathbf v}^{(1)}\cdot {\mathbf v}^{(2)}, \;\; S= s \sqrt{(1-|{\mathbf v}^{(1)}|^2) (1-|{\mathbf v}^{(2)}|^2)}.
\end{eqnarray}
}
{ We remark that the norm of the vectors ${\mathbf v}^{(j)}$, $j=1,2$, defined in (\ref{vectors})  is less than one (as mentioned in the Introduction) and $a^{(j)}_{0}$, $j=1,2$, can not equal zero (otherwise $S=0$), as noted in Eq.(\ref{pureK}).}
It is simple to show that in this case 
\begin{eqnarray}\label{smax}
s=1/\sqrt{K}. 
\end{eqnarray}
{In fact,  Eq.(\ref{sconstr}) means that there are $K$ nonzero elements in the selected column of $W^+$ and each of them  equals $s$. Then Eq.(\ref{smax}) 
follows from the normalization condition for this column.}
This is the maximal value of the scale coefficient $s$ in our
protocol, $s$ decreases with an increase in the length of the channels
as 
will be seen
 in {the example of Sec.\ref{Section:Example2}.}

The final step of the protocol is the measurement of the intensity
$I_2$ of the second-order coherence, given by the second-order
coherence matrices $\rho^{(R; \pm2)}$ of the receiver:
\begin{eqnarray}\label{I2}
I_2={\mbox{Tr}}\rho^{(R;2)}\rho^{(R;-2)} = (\rho^{(R)}_{00;11})^2=
{
S^2 ({\mathbf v}^{(1)} \cdot {\mathbf v}^{(2)})^2.
}
\end{eqnarray}
Thus the measured quantity $I_2$ 
is proportional to the square of the   scalar
product of the vectors ${\mathbf v}^{(1)}$ and ${\mathbf v}^{(2)}$.


{\subsection{Generalization to complex vectors ${\mathbf v}^{(i)}$}
\label{Section:complex}
The reason  requiring the reality of $ {\mathbf v}^{(i)}$ is in the
structure of the matrix element $\rho^{(R)}_{00;11}$ given in
(\ref{sp}). For the scalar product of complex vectors ${\mathbf
  v}^{(i)}$ this formula 
 should be replaced by
\begin{eqnarray}\label{spcompl}
\rho^{(R)}_{00;11} =(a^{(1)}_{0})^* a^{(2)}_{0}  s \sum_{i=1}^K a^{(1)}_{i} \Big(a^{(2)}_{i}\Big)^*,
\end{eqnarray}
which can be achieved
 by a simple unitary transformation of the initial
state of the {first} sender. This transformation must transfer the
amplitude  {$a^{(1)}_0$} in the state 
{$|\psi_1\rangle$} (\ref{pureK}) to a state with  two excited spins,
say, the 1st and 2nd, $| 12 \rangle$:
{
\begin{eqnarray}
U:\;\;  a^{(1)}_0|0\rangle + \sum_{n=1}^K a^{(1)}_{n} |{K-n+1}\rangle \;\; \rightarrow \;\; \sum_{n=1}^K a^{(1)}_{n} |{K-n+1}\rangle + a^{(1)}_0|12\rangle.
\end{eqnarray}
}
In terms of the MQ-coherence matrices{,} this transformation means 
transferring
 the elements of the 
{ $-1$st-order}
 coherence matrix to the 
{$1$st-order}
 coherence matrix, while the elements of the 
{$1$st-order}
coherence matrix
 of the state {$|\psi_1\rangle$} become zeros. 
There is no principal 
difficulty in constructing the appropriate unitary transformation of
the extended receiver in this case, but 
the spin dynamics 
 involved in the time evolution then
must be
extended to the three-excitation subspace. We 
will not study the scalar product of complex vectors  in more detail.}

\subsection{Example: scalar product of two-element vectors, $K=2$}
\label{Section:Example1}
We illustrate our protocol for the simplest case of two-dimensional
vectors. 
The unitary transformation in this case is block-diagonal
\begin{eqnarray}\label{EW}
&&W^+={\mbox{diag}}(1,W_1,W_2),\\\label{W2}
&&
W_2=
\left(
\begin{array}{cccccc}
 -1 & 0 & 0 & 0 & 0 & 0 \\
 0 & 0 & 1/\sqrt{2} & 0 &-1/\sqrt{2} & 0 \\
 0 & 1/\sqrt{2} & 0 & -1/\sqrt{2} & 0 & 0 \\
 0 & 1/\sqrt{2} & 0 & 1/\sqrt{2} & 0 & 0 \\
 0 & 0 &1/\sqrt{2}& 0 & 1/\sqrt{2} & 0 \\
 0 & 0 & 0 & 0 & 0 & 1
\end{array}
\right),
\end{eqnarray}
where the {two excitation} block $W_2$ refers to the  basis of the {extended} receiver states
\begin{eqnarray}
{|0011\rangle,\;\;|0101\rangle,\;\;|0110\rangle,\;\;|1001\rangle,\;\;|1010\rangle,\;\;|1100\rangle.}
\end{eqnarray}
In (\ref{EW}),
$W_1$ is the $4\times 4$ block in the one-excitation subspace,
 which is not important in our case.
Eq.(\ref{smax}) yields
$s=\frac{1}{\sqrt{2}}$. {We note that, according to
  eq.(\ref{rhoR}), only the third column of $W_2$ in (\ref{W2}) is
  important for the scalar product. 
 The other columns  only serve to fulfill}
 the unitarity condition for $W_2$. 

{{}
From the point of view of quantum information processing, it
is important to note  that the block $W_2$ in the form (\ref{W2}) can
be written in terms of standard quantum gates, namely { the} one-qubit
rotations and the two-qubit 
{
CNOT} gate.
We denote by  $C_{ij}$  
{
the CNOT {entangling the $i$th and $j$th spins}
with {the} control qubit $i$  written
 {in the basis $|0\rangle$, $|j\rangle$, $|i\rangle$, $|ji\rangle$:}}
\begin{eqnarray}
C_{ij}=\left(
\begin{array}{cccc}
1&0&0&0\cr
0&1&0&0\cr
0&0&0&1\cr
0&0&1&0
\end{array}
\right).
\end{eqnarray}
Let $R_{yi}(\beta)$ be the $y$-rotation of the $i$th qubit:
\begin{eqnarray}\label{Ry}
R_{yi}(\beta)=e^{i \beta I_{yi}}=\left(
\begin{array}{cc}
\cos\frac{\beta}{2}& \sin \frac{\beta}{2}\cr
-\sin\frac{\beta}{2}&\cos\frac{\beta}{2}
\end{array}
\right).
\end{eqnarray}
We can introduce a two-qubit operation with the qubits $i$ and $j$ which  commutes with $I_{zi}+I_{zj}$:
\begin{eqnarray}\label{E1}
E_{ij}(\beta) = C_{ij} R_{yi}(\beta) C_{ji} R_{yi}(-\beta)C_{ij}.
\end{eqnarray}
Then the unitary transformation
\begin{eqnarray}\label{UER1}
W^+ = E_{12}(0) E_{34}(\frac{\pi}{4}) E_{23}(-\frac{\pi}{2})  E_{34}(\frac{\pi}{4})E_{12}(\frac{\pi}{4})
\end{eqnarray}
produces the block $W_2$ (\ref{W2}) in the two-excitation subspace.}



\section{Scalar product of two remote vectors, Fig.\ref{Fig:CL1} }
\label{Section:ScPr}
\label{Section:GP}
In this section we consider the general situation shown in
Fig.\ref{Fig:CL1}, where  the $K$-qubit senders $S_1$ and
$S_2$ are connected to the 2-qubit receiver $R$ 
by transmission lines of{,  in general, different lengths}.
We construct the general protocol for calculating the scalar product of $K$-dimensional vectors ($K>1$) with real elements (\ref{vectors}).
To this end we consider the  pure 
states (\ref{pureK}) of  two $K$-qubit senders each containing at
most
 a single excitation
and with   real coefficients $a^{(i)}_k$.
The initial state of the 
{ entire}
 quantum system reads:
\begin{eqnarray}\label{inst}
\rho(0)=\rho^{(S_1)}(0)\otimes \rho^{(TL_1,R,TL_2)}(0) \otimes \rho^{(S_2)}(0),\;\;{
\rho^{(S_i)}(0) = |\psi_i\rangle \langle \psi_i|, \;\;i=1,2},
\end{eqnarray}
where $|\psi_i\rangle$, $i=1,2$, are  the sender's pure states (\ref{pureK}) and
{
$\rho^{(TL_1,R,TL2)}(0)$} 
 is the initial state of the 
remainder of the  spin chain. 

We show,
 similar to Sec.\ref{Section:ScalarProdD2},
 that the scalar product of two vectors appears in the only element of the second-order coherence matrix of the two-qubit receiver.}

We consider  
{
the unitary transformation 
$W$ defined in (\ref{W}) and  define the multi-indices $I_1$, $\dots$,
$I_5$ related with, respectively, subsystems $S_1$, $TL_1$, $R$,
$TL_2$  and $S_2$.} 
Each multi-index $I_i$ of a {$k_i$}-qubit subsystem consists of 
a set of {$k_i$} zeros and ones where the one
 at the $k$th position corresponds to the excited $k$th spin.
Then, using the initial state (\ref{inst}) 
{ and time evolution included in} 
the unitary operator $W$ (\ref{W}), we obtain,
in a way analogous to (\ref{eq:10})
\begin{eqnarray}
&&
\rho^{(R)} = {\mbox{Tr}}_{S_1,TL_1,TL_2,S_2}\,(\rho) \;\;\Rightarrow \\\label{rho1}
&&
\rho^{(R)}_{N_3;M_3} = \sum_{N_1,N_2,N_4,N_5,\{I\},\{J\}} 
W_{{\{N\}; \{I\}}} 
\rho^{(S_1)}_{I_1;J_1}
\rho^{(TL_1,R,TL_2)}_{I_2I_3I_4;J_2J_3J_4}
\rho^{(S_2)}_{I_5;J_5}W^+_{{\{J\}}; N_1 N_2M_3N_4N_5},
\end{eqnarray}
where $\{I\}$ is short for $(I_1,I_2,I_3,I_4,I_5)$ and $\{J\}${, $\{N\}$}
analogously. Note that we have suppressed the time dependence of $W$
for simplicity. The density matrices on the rhs are to be taken at
$t=0$, while $\rho^{(R)}$ depends on time.
We can reasonably assume the system to be initialized such that at
$t=0$ excitations are only present in the senders. The initial density
operator of the connection between $S_1$ and $S_2$ then is
\begin{eqnarray}\label{TLRTL}
\rho^{(TL_1,R,TL_2)}= |0_20_30_4\rangle \langle 0_20_30_4|,
\end{eqnarray}
and its
 only  nonzero element is 
$\rho^{(TL_1,R,TL_2)}_{0_20_30_4;0_20_30_4}=1$, where $0_i$ is the zero value of the multi-index, associated with the $i$th subsystem. Then  expression (\ref{rho1}) becomes simpler:
\begin{eqnarray}\label{rho2}
\rho^{(R)}_{N_3;M_3} = \sum_{N_1,N_2,N_4,N_5} \sum_{I_1,I_5,J_1,J_5}
W_{{\{N\}}; I_1 0_20_30_4I_5}
\rho^{(S_1)}_{I_1;J_1}\rho^{(S_2)}_{I_5;J_5}W^+_{J_1 0_20_30_4J_5; N_1 N_2M_3N_4N_5}.
\end{eqnarray}
Now we consider the second-order coherence matrix which consists of a single element with 
 $N_3=\{0,0\}$, $M_3=\{1,1\}$, and take into account that we stay in
 the
subspace with two excitations at most, and that $W_{0_1 0_2 00 0_40_5;
  0_1 0_20_30_40_5}=1$. {
By an argument similar to that leading from
(\ref{eq:10}) to (\ref{rhoR}), } 
 (\ref{rho2}) reduces to the following form:
\begin{eqnarray}\label{red3}
\rho^{(R)}_{00;11} =  
\sum_{|J_1|=1,|J_5|=1}
\rho^{(S_1)}_{0_1;J_1}\rho^{(S_2)}_{0_5;J_5}W^+_{J_1 0_20_30_4J_5; 0_1 0_2 11 0_40_5}.
\end{eqnarray}
Using the free parameters of the unitary transformation $U^{(ER)}$ we
can  
achieve
\begin{eqnarray}\label{syst1}
W^+_{J_1 0_20_30_4J_5; 0_1 0_2 11 0_40_5} =0,\;\; J_1 \neq J_5.
\end{eqnarray}
There are $K(K-1)$ equations in this system. 
If we also satisfy the $K$ equations 
\begin{eqnarray}\label{syst2}
W^+_{J_1 0_20_30_4J_1; 0_1 0_2 11 0_40_5} =s=const,\;\; |J_1|=1,
\end{eqnarray}
then (\ref{red3}) gets the form ($0_5\equiv 0_1$)
\begin{eqnarray}\label{red32}
\rho^{(R)}_{00;11} = s \sum_{i=1}^K
\rho^{(S_1)}_{0_1;J_1^{(i)}}\rho^{(S_2)}_{0_1;{J_5}^{(i)}} = s {\mathbf v}^{(1)}\cdot {\mathbf v}^{(2)},
\end{eqnarray}
where 
the vectors ${\mathbf v}^{(i)}$,
 $i=1,2$, are defined in (\ref{vectors}) and (\ref{star}).

In order to construct the required  element $\rho^{(R)}_{00;11}$ of
the two-qubit receiver's density matrix  the unitary transformation
$W$ has to satisfy conditions (\ref{syst1}) and (\ref{syst2}). These
conditions involve only one  column of  $W^+$, { namely the one with the index $(0_1 0_2 11 0_40_5)$. This column corresponds to 
the column with the index $(\underbrace{0\dots01}_{K_{ 1}}
\underbrace{10\dots0}_{K_{ 2}})$  in $(U^{(ER)})^+$,}
similar to Sec.\ref{Section:ScalarProdD2}. The
ones in the above multi-index refer to the nodes of the receiver. {We denote this column $(U^{(ER)})^+_{\underbrace{0\dots01}_{K_{ 1}}
\underbrace{10\dots0}_{K_{ 2}}}$.} 
Due to the conservation law (\ref{constr}), the nonzero elements of
the column are those that connect to other two-excitation states of
the extended receiver. The dimension of the space spanned by these
states, and hence, the number of nonzero elements in the column under
discussion, is  $P = \frac 12 {N^{(ER)}(N^{(ER)}-1)}$, where we
temporarily denote  the number of qubits in the extended receiver by
{$N^{(ER)}=K_1+K_2$}. Due to unitarity the nonzero elements are points on the
$P$-dimensional unit sphere
and can be parametrized in terms of $2P-1$ angles:
\begin{eqnarray}\label{vpar}{
(U^{(ER)})^+_{\underbrace{0\dots01}_{K_{ 1}}
\underbrace{10\dots0}_{K_{ 1}}}=}\left(
\begin{array}{c}
e^{i\varphi_1}\sin \alpha_1 \sin \alpha_2 \dots \sin\alpha_{P-1},\cr
e^{i\varphi_2}\cos \alpha_1 \sin \alpha_2 \dots \sin\alpha_{P-1},\cr
e^{i\varphi_3}\cos \alpha_2 \sin \alpha_3 \dots \sin\alpha_{P-1},\cr
\cdots \cr
e^{i\varphi_{P}}\cos\alpha_{P-1}
\end{array}
\right).
\end{eqnarray}
The remaining
 columns of the unitary transformation can be constructed to satisfy
 the unitarity condition $U^{(ER)} (U^{(ER)})^+=E_{ER}$,
where $E_{ER}$ is the unit operator on the extended receiver. 
Thus, we have $2P-1$ real parameters $\alpha_i$ and $\varphi_i$ to
satisfy the  $K^2$ complex equations (\ref{syst1}) and
(\ref{syst2}). The condition
\begin{eqnarray}\label{PK}
2P-1=N^{(ER)}(N^{(ER)}-1)-1 \ge 2 K^2\;\;{\Rightarrow \;\; N^{(ER)}\ge \frac{1}{2}(1+\sqrt{5+ 8 K^2}}
&&
\end{eqnarray}
defines the minimal size of the extended receiver. {In particular, the choice
$K_1=K_2=K$} obviously fulfills this condition for all $K \ge 2$.

\subsection{Example: scalar product of  two-element vectors using two  channels of 20 nodes}
\label{Section:Example2}

We consider the two-channel communication line governed by the  $XX$-Hamiltonian (\ref{XX}).  Each channel consists of 20 nodes ($N=40$) with two pairs of coupling constants  adjusted for the high probability state transfer between the end nodes of an isolated channel \cite{SZ2017} (see Fig.\ref{Fig:CL1}):
\begin{eqnarray}
&&
D_1=D_{N/2-1}=D_{N/2+1}=D_{N-1}=0.55,\\\nonumber
&&
D_2=D_{N/2-2}=D_{N/2+2}=D_{N-2}=0.817.
\end{eqnarray}
The coupling 
{
 between the two channels is weak, the coupling constant being
 $D_{N}=0.006$.
}  
In this case the evolution operator 
$V(t)$ is essential in $W(t)$ {(see Eq.(\ref{W}))}
 and there is an  
optimal time instant providing the maximal value for the parameter $s$ in (\ref{red32}). 

As an example, we consider
two-qubit senders $i=1,2$ encoding  two-element real vectors ${\mathbf v}^{(i)}$. This corresponds to $K=2$ in (\ref{pureK}) and in (\ref{vectors}). 
We use a four-qubit extended receiver in the communication line, {setting $K_1=K_2=2$}. 
Then $P=6$, so that we  have 11 real parameters in (\ref{vpar}) to satisfy 4 complex  equations (\ref{syst1}), (\ref{syst2}). 
The optimization yields that the maximal $s=0.6813$ is achieved at 
$t=26.441$ (this  time instance coincides with that for the
high-probability state transfer \cite{SZ_2016} between the end nodes
of a
single
 channel) with the parameters in (\ref{vpar})
\begin{eqnarray}\label{alpphi}
&&\alpha_1=3.135160, \;\;\alpha_2=1.570857, \;\;\alpha_3=4.712397, \;\;\alpha_4=5.497855, \;\;\alpha_5=1.581785,\\\nonumber
&&
\varphi_1= 5.526328, \;\; \varphi_2=0.000065 , \;\; \varphi_3=1.497402 , \;\; \varphi_4=0.999731 , \;\; \varphi_5=3.141532, \\\nonumber
&&\varphi_6= 1.319482.
\end{eqnarray}
{We note that the above calculated $s$ is less than the one obtained 
{ from formula (\ref{smax})}
 in Sec.\ref{Section:Example1}.} { In addition, the values of $\alpha_i$, $i=1,\dots,5$, and $\varphi_2$, $\varphi_5$  are close to  multiples of $\pi/4$:
\begin{eqnarray}\label{alpphi0}
&& \alpha_1\approx\pi, \alpha_2\approx\frac{\pi}{2}, \alpha_{3}\approx\frac{3 \pi}{2}, \alpha_{4}\approx\frac{7\pi}{4},
\alpha_5\approx\frac{\pi}{2}, 
\varphi_2\approx0, \varphi_5\approx\pi.
\end{eqnarray}
 This is not by  chance. With these $\alpha_i$ and $\varphi_i$, the column (\ref{vpar}) only by sign differs from the third column of the matrix $W_2$ in (\ref{W2})
 (which is responsible for the scalar product in the ''senders only''
 scheme). 
 This change in sign as well as the deviation of parameters (\ref{alpphi}) from values (\ref{alpphi0})
is due to  the evolution and imperfection of state transfer.}
 
 {{}
Again, the constructed unitary transformation can be generated by the set of CNOTs and one-qubit rotations as follows. 
Let $C_{ij}$  be CNOT  with control qubit
 $i$, $R_{yi}(\beta)$ be the $y$-rotation (\ref{Ry}) and $R_{zi}(\beta)$ be the $z$-rotation  of the $i$th spin:
\begin{eqnarray}\label{Rz}
R_{zi}(\beta)=e^{i \beta I_{zi}}=\left(
\begin{array}{cc}
e^{i\frac{\beta}{2}}& 0\cr
0&e^{-i\frac{\beta}{2}}
\end{array}
\right).
\end{eqnarray}
We can introduce the two-qubit operation on the qubits $i$ and $j$ which  commutes with $I_{zi}+I_{zj}$:
\begin{eqnarray}\label{E2}
E_{ij}(\alpha,\beta) = C_{ij}R_{zi}(\alpha) R_{yi}(\beta) R_{zi}(-\alpha) C_{ji} R_{zi}(\alpha)R_{yi}(-\beta)R_{zi}(-\alpha)C_{ij}.
\end{eqnarray}
Then the unitary transformation $U^{ER}$ with the values  (\ref{alpphi}) for the parameters in (\ref{vpar}) can be represented, for instance, in the following form:
\begin{eqnarray}\label{UER2}
(U^{ER})^+ = E_{12}(\alpha_1,\beta_1) E_{23}(\alpha_2,\beta_2) E_{34}(\alpha_3,\beta_3)  E_{43}(\alpha_4,\beta_4) E_{32}(\alpha_5,\beta_5) E_{21}(\alpha_6,\beta_6),
\end{eqnarray}
where, (with the accuracy $\sim 10^{-6}$),
\begin{eqnarray}
&&
\alpha_{1} =0.056765, \;\beta_{1} =5.496129,\; \alpha_{2} =6.276980, \;
 \beta_{2} =1.577448,\;\;
 \\\nonumber
 &&
 \alpha_{3} =6.134857,\; \beta_{3} =0.320085, \;
 \alpha_{4} =6.184914, \;\beta_{4} =0.471440,
 \\\nonumber
 &&
 \alpha_{5} =6.263752, \;
 \beta_{5} =1.562174,\; \alpha_{6} =6.226368,\; \beta_{6} =0.786962.
\end{eqnarray}

It is interesting to note
 that the
constraints (\ref{sconstr}) for the four-node chain ($K=2$) can be also satisfied with the unitary transformation $U^{(ER)}$ having the structure (\ref{UER2}):
\begin{eqnarray}\label{UER12}
&&
W^+=(U^{ER})^+ =\\\nonumber
&&E_{12}(0, 2 \varphi) E_{23}(0,4 \varphi) E_{34}(0,-\varphi)  E_{43}(0,-\varphi) E_{32}(0,4 \varphi) E_{21}(0,-2 \varphi), \;\;\varphi=\frac{\pi}{8}.
\end{eqnarray}
We do not present  the explicit formula for the block $W_2$ in this case. It differs from (\ref{W2}), but the  third column of this block, which controls the constraints (\ref{sconstr}), is the same. }

\section{Conclusion}
\label{Section:conclusions}

Using a local unitary transformation of the so-called extended receiver we obtain a scalar product of two real vectors and place the result in
the  element of  the second-order MQ-coherence matrix  of the
receiver. These vectors are initially encoded in the pure states of
two senders which are, generically, remote from the receiver. Thus,
the encoded vectors  evolve along the spin-1/2 channels to the
receiver, therefore   getting mixed. The unitary transformation at the
extended receiver ({which includes $K_1$ and $K_2$ qubits from, respectively, the first and second   channels}) is
used  to remove extra terms in the resulting expression for the above
element, so that the remaining terms form an expression proportional
to the scalar product of the original vectors. The {factor $s$ in the  proportionality
coefficient $S$ (\ref{S})} can not exceed $1/\sqrt{K}$, which is found for the scheme
without
finite-length transmission lines
 $TL_i$, Fig.\ref{Fig:CL2},  and decreases  with an increase in the
 channel lengths. {{} For the simplest example of the scalar product of
 two-element vectors ($4$-node extended receiver),  we show that the
 unitary transformation $U^{ER}$ can be represented as a 
combination of CNOTs
 and one-qubit rotations, which is important for programming the scalar product on quantum computers. }

The dimensionality of the quantum  system  used for implementing the
scalar-multiplication protocol
 does not depend on the required accuracy of calculations, but only
 on
 the dimensionality of the original vectors and on the distance
between the senders and the receiver (the lengths of $TL_i$,
$i=1,2$). In addition, the result of 
the multiplication is transferred to a particular element of the
receiver's density matrix without performing measurements on any
particular subsystem. Therefore, the protocol is  completely quantum
and does not  involve any classical 
{ step}
 except
{ for} 
the initialization of the vectors $\mathbf v^{(i)}$. Hence, the obtained scalar product can be used in further quantum calculations.  

The derived unitary transformation of the extended receiver  depends
on the parameters of the communication line, on the Hamiltonian
governing the quantum evolution and on the length $K$ of the vectors
to be multiplied. 
Once
 constructed, this transformation can be used for multiplying any pair of  $K$-dimensional vectors.
 
We 
wish to
 emphasize that the proposed protocol allows to multiply  vectors with
 real elements and is based on 
 unitary transformations (the time evolution operator and the local
 unitary transformation $U^{(ER)}$ on the extended receiver) which
 conserve
 the excitation number in the spin system and therefore do not mix coherence matrices of different orders. { However, the  generalization  to  complex vectors is quite straightforward and was outlined in Sec.\ref{Section:complex}.}

{This work performed in accordance with the state task, state registration No. 0089-2019-0002. }
 One of the authors (AZ) acknowledges the support from Presidium of RAS, Program No.5 ''Photonic technologies in probing inhomogeneous media and biological objects''.


\end{document}